\documentclass{IEEEtran}
\usepackage{cite}
\usepackage{amsmath,amssymb,amsthm,amsfonts}
\usepackage{algorithm,algorithmic}
\usepackage{graphicx}
\usepackage{textcomp}

\newtheorem{theorem}{Theorem}

\newtheorem{remark}{Remark}
\newtheorem{assumption}{Assumption}

\def\BibTeX{{\rm B\kern-.05em{\sc i\kern-.025em b}\kern-.08em
    T\kern-.1667em\lower.7ex\hbox{E}\kern-.125emX}}

\title{Learning-Based Robust Bayesian Persuasion with Conformal Prediction Guarantees}

\author{Heeseung Bang, \IEEEmembership{Member, IEEE}, and Andreas A. Malikopoulos, \IEEEmembership{Senior Member, IEEE}
    \thanks{This research was supported in part by NSF under Grants CNS-2401007, CMMI-2348381, IIS-2415478, and in part by MathWorks.}
	\thanks{Heeseung Bang is with the School of Civil and Environmental Engineering, Cornell University, Ithaca, NY 14850, USA. {\tt\small email: \{h.bang\}@cornell.edu}}
    \thanks{Andreas Malikopoulos is with the School of Civil and Environmental Engineering and Systems Engineering Program, Cornell University, Ithaca, NY 14850 USA. {\tt\small email: \{amaliko\}@cornell.edu}}
}

\begin{document}

\maketitle

\begin{abstract}
Classical Bayesian persuasion assumes that senders fully understand how receivers form beliefs and make decisions—an assumption that rarely holds when receivers possess private information or exhibit non-Bayesian behavior. In this paper, we develop a learning-based framework that integrates neural networks with conformal prediction to achieve \emph{robust persuasion} under uncertainty about receiver belief formation. The proposed neural architecture learns end-to-end mappings from receiver observations and sender signals to action predictions, eliminating the need to identify belief mechanisms explicitly. Conformal prediction constructs finite-sample valid prediction sets with provable marginal coverage, enabling principled, distribution-free robust optimization. We establish exact coverage guarantees for the data-generating policy and derive bounds on coverage degradation under policy shifts. Furthermore, we provide neural network approximation and estimation error bounds, with sample complexity $O(d \log(|\mathcal{U}||\mathcal{Y}||\mathcal{S}|)/\varepsilon^2)$, where $d$ denotes the effective network dimension, and finite-sample lower bounds on the sender’s expected utility. Numerical experiments on smart-grid energy management illustrate the framework’s robustness.
\end{abstract}


\section{Introduction}
\label{sec:introduction}

Bayesian persuasion examines the strategic transmission of information between an informed sender and a decision-making receiver. In the classical formulation~\cite{kamenica2011bayesian}, the sender observes a private state, commits to a signaling policy, and the receiver updates beliefs using Bayes’ rule to select an optimal action. While the original framework was developed in the context of economic markets~\cite{hanson2003combinatorial}, its underlying principles extend naturally to a broader range of information-sharing scenarios, including governance mechanisms and interactive computational systems~\cite{venkatesh2023connected,sun2025recommendation}.

The classical persuasion model assumes that the sender possesses complete knowledge of the receiver’s decision-making process and that the receiver performs fully Bayesian updates of beliefs under known prior distributions. Recent research has sought to relax these assumptions, addressing scenarios where senders face uncertainty about receiver preferences, belief formation, or informational constraints.
For example, Castiglioni et al. \cite{castiglioni2020online} addressed unknown receiver utilities using no-regret algorithms and extended to multi-receiver settings \cite{castiglioni2021multi}, while Bernasconi et al.\cite{bernasconi2022sequential} tackled sequential information design without knowing prior distributions and considered farsighted receivers \cite{bernasconi2023persuading}.
Other efforts have addressed dynamic environments \cite{renault2017optimal,farhadi2022dynamic,massicot2025almost,venkatesh2025persuasion} and implementation challenges \cite{sayin2021bayesian}.
In dynamic Markovian settings, Lehrer and Shaiderman \cite{lehrer2022markovian} characterized optimal sender payoffs, \cite{iyer2023markov} extended this to history-dependent beliefs, and Bacchiocchi et al. \cite{bacchiocchi2024markov} established regret guarantees when senders lack environmental knowledge.
Despite these advances, most existing approaches continue to assume that receivers adhere to Bayesian principles and possess knowledge of the underlying prior distributions. In practice, however, receivers often update their beliefs through non-Bayesian mechanisms~\cite{gennaioli2010comes}, influenced by cognitive limitations, bounded rationality, and prior experiences. This discrepancy raises a fundamental question: how can a sender design effective persuasion strategies when the receiver’s belief formation process is uncertain?

Traditional robust optimization approaches for mechanism design \cite{chremos2022CSMArticle}, such as worst-case optimization over uncertainty sets \cite{ben2009robust} and distributionally robust optimization using Wasserstein balls \cite{kuhn2019wasserstein}, face several challenges in the persuasion setting. Worst-case optimization can be overly conservative, leading to significant utility loss. Distributionally robust methods require carefully specified uncertainty sets that are difficult to construct without strong prior knowledge \cite{venkatesh2024stochastic}. Moreover, the computational complexity of nested minimax problems scales poorly with the dimensions of the state and action spaces. Existing learning approaches for Bayesian persuasion \cite{castiglioni2020online,bernasconi2022sequential} either assume known belief mechanisms or lack finite-sample statistical guarantees on coverage and utility.

This paper addresses these challenges through a distributionally robust framework that integrates conformal prediction with neural networks. 
First, we show that identifying receiver priors separately from Bayesian updating mechanisms is fundamentally infeasible from observed data, since actions are observed only after receivers process both private observations and sender signals. To overcome this, our neural network learns the composite mapping from observations, signals, and policies to actions—sufficient for policy optimization and free from the identification problem. 
Second, we construct conformal prediction sets that provide finite-sample valid coverage for receiver actions. For data collected under a baseline policy, we prove exact marginal coverage and derive bounds on coverage degradation under new policies using three shift measures: joint distribution shift, conditional mechanism shift, and calibration error. This enables single-policy learning where data from one policy supports robust optimization over alternatives. 
Third, we establish comprehensive performance guarantees, including neural network approximation and estimation error bounds, a sample complexity of $O((d\log(|\mathcal{U}||\mathcal{Y}||\mathcal{S}|/\varepsilon)+\log(1/\delta))/\varepsilon^2)$ for $\varepsilon$-optimal policy selection (where $d$ is the effective network dimension), and finite-sample lower bounds on sender utility. The framework transforms robust policy selection into standard optimization over conformal sets, providing explicit probabilistic guarantees without nested minimax computation.

Our framework offers several practical advantages. Neural networks capture complex belief-formation processes with computational tractability, while conformal prediction provides distribution-free guarantees through simple quantile computations. The approach scales efficiently to high-dimensional settings using standard training procedures. We demonstrate its effectiveness through numerical experiments in realistic smart-grid demand-response scenarios with private receiver information.

The remainder of the paper proceeds as follows. In Section~II, we formulate the problem of robust Bayesian persuasion with private receiver observations. In Section~III, we develop the conformal prediction framework used to construct finite-sample valid uncertainty sets for receiver actions. In Section~IV, we present the main theoretical results, including coverage bounds under policy shift, neural network performance guarantees, and finite-sample utility bounds.
In Section~V, we demonstrate the effectiveness of the proposed approach through numerical experiments in a smart-grid demand-response scenario. Finally, in Section~VI, we conclude the paper and outline directions for future research.

\section{Problem Formulation}
\label{sec:problem_formulation}

Throughout this paper, we denote random variables with upper-case letters (e.g., $X$, $Y$, $S$, $U$) and their realizations with lower-case letters (e.g., $x$, $y$, $s$, $u$).

\subsection{Standard Bayesian Persuasion}

Consider the strategic information sharing between a sender and a receiver. The sender privately observes state $x \in \mathcal{X}$ sampled from prior $\mu \in \Delta(\mathcal{X})$, where $\Delta(\mathcal{X})$ denotes the probability simplex over $\mathcal{X}$. Based on a pre-committed signaling policy $\pi: \mathcal{X} \to \Delta(\mathcal{S})$, the sender selects signal $s \in \mathcal{S}$ with probability $\pi(s|x)$.
Upon receiving signal $s$, the receiver computes the posterior using Bayes' rule as
\begin{equation}
p^\pi(x|s) = \frac{\pi(s|x)\mu(x)}{\sum_{x'} \pi(s|x')\mu(x')},
\end{equation}
and takes action $u = g(s) = \arg\max_u \sum_x p^\pi(x|s) r_r(x, u)$ where $r_r: \mathcal{X} \times \mathcal{U} \to \mathbb{R}$ is the receiver's reward function.

The sender anticipates the receiver's response and chooses the policy maximizing their expected reward as
\begin{equation}
\hat{\pi} = \arg\max_\pi \sum_x \mu(x) \sum_s \pi(s|x) r_s(x, g_\pi(s)),
\end{equation}
where $r_s: \mathcal{X} \times \mathcal{U} \to \mathbb{R}$ is the sender's reward function.

\subsection{Robust Bayesian Persuasion with Private Observations}

We extend this framework to account for receivers with private observations and uncertain belief formation. Consider finite state space $\mathcal{X}$, observation space $\mathcal{Y}$, signal space $\mathcal{S}$, and action space $\mathcal{U}$.
The sender and receiver follow the interaction protocol described as follows. Nature draws state $x \sim \mu_X$ and receiver's private observation $y \sim \mu_{Y|X}(\cdot|x)$. The sender observes $x$ and sends signal $s \sim \pi(\cdot|x)$. The receiver observes $(y, s)$ and chooses action $u$.

The receiver selects the best action as
\begin{equation}
u = g(y, s) = \arg\max_u \sum_{x \in \mathcal{X}} p_r^\pi(x|y, s) r_r(x, u),
\end{equation}
where $p_r^\pi(x|y, s)$ represents the receiver's posterior belief over states after observing both their private observation $y$ and the sender's signal $s$.

In the classical Bayesian persuasion setting without private observations, the receiver knows the prior distribution $\mu_X$ and can directly apply Bayes’ rule upon receiving a signal $s$. When private observations are present, however, the receiver first forms an initial belief based on their observation $y$ before observing $s$. We formalize this through a belief function $\theta^*: \mathcal{Y} \to \Delta(\mathcal{X})$, where $\theta^*(y)(x) = p_r(x|y)p_r(y)$ denotes the receiver’s belief over states after observing $y$ but prior to receiving the signal. This belief function may deviate from the Bayesian posterior $p(x|y) \propto \mu_{Y|X}(y|x)\mu_X(x)/\mu_Y(y)$ due to cognitive limitations, heuristic reasoning, or incomplete knowledge of the joint distribution $(\mu_X, \mu_{Y|X})$. Upon receiving the signal $s$, the receiver updates this belief using Bayes’ rule,
\begin{equation}
p_r^\pi(x|y, s) = \frac{\pi(s|x)\theta^*(y)(x)}{\sum_{x' \in \mathcal{X}} \pi(s|x')\theta^*(y)(x')}.
\end{equation}

The key challenge is that the sender does not know the receiver’s belief function $\theta^*$. Even if the sender is aware of $\mu_{Y|X}$, inferring how receivers form beliefs from their observations remains difficult when they rely on heuristics, approximations, or are influenced by factors beyond the statistical model. Consequently, the sender faces a robust optimization problem that requires learning from data about receiver responses while accounting for uncertainty in belief formation.

We formulate the sender’s problem as a robust optimization,
\begin{equation}
\max_{\pi \in \Pi} \sum_{x \in \mathcal{X}} \mu_X(x) \sum_{y \in \mathcal{Y}} \mu_{Y|X}(y|x)  \sum_{s \in \mathcal{S}} \pi(s|x)\min_{u \in \mathcal{U}(y,s,\pi)} r_s(x, u),
\label{eq:robust_opt}
\end{equation}
where $\mathcal{U}(y, s, \pi)$ denotes the uncertainty set of plausible receiver actions given observations $(y, s)$ under policy $\pi$. This formulation optimizes the sender’s expected utility against worst-case receiver responses within these sets, thereby eliminating the need to explicitly model the receiver’s belief formation mechanism.

The key challenge is constructing $\mathcal{U}(y, s, \pi)$ from finite data with statistical guarantees. Our approach uses conformal prediction to construct sets $C_{1-\alpha}(y, s, \pi)$ such that
\begin{equation}
\mathbb{P}\{U_{\text{true}} \in C_{1-\alpha}(Y, S, \pi)\} \geq 1 - \alpha,
\end{equation}
where $U_{\text{true}}$ is the receiver's actual action and $1-\alpha$ is confidence level. This transforms the problem into
\begin{equation}
\max_{\pi \in \Pi} \mathbb{E}_{X,Y,S}^\pi\left[\min_{u \in C_{1-\alpha}(Y,S,\pi)} r_s(X, u)\right].
\label{eq:conformal_robust_opt}
\end{equation}

\subsection{Neural Architecture for End-to-End Action Prediction}

A fundamental identification challenge arises from observational data. We observe tuples $(x_i, y_i, s_i, \pi_i, u_i)$, where the receiver has observed the private signal $y_i$, the sender’s signal $s_i$, and the signaling policy $\pi_i$ before selecting action $u_i$. However, we never observe actions taken based solely on $y_i$ before exposure to $s_i$ and $\pi_i$, nor do we observe the receiver’s beliefs directly.  

Consequently, it is impossible to separately identify the prior belief function $\theta^*(y)$ from the Bayesian updating mechanism, since the action $u$ depends on the posterior $p_r(x|y,s)$. Any attempt to infer $\theta^*(y)$ from actions observed after $(y,s)$ conflates the receiver’s prior belief with the influence of the signal. This introduces a form of confounding, as the signal $s$ simultaneously affects both the belief update and the observed action.

To address this challenge, we directly learn the composite mapping from $(y, s, \pi)$ to receiver actions rather than attempting separate identification. This approach is sufficient because the sender's utility $r_s(x, u)$ only depends on the realized state and action, and conformal prediction constructs uncertainty sets over actions $u$ rather than beliefs. By learning the end-to-end mapping, we bypass the identification problem while retaining all information necessary for policy optimization.

We define a neural network  
\begin{equation}
f_\theta: \mathcal{Y} \times \mathcal{S} \times \Pi \to \Delta(\mathcal{U}),
\end{equation}
that maps the receiver’s observation, the sender’s signal, and the signaling policy to a probability distribution over actions, where $f_\theta(y, s, \pi)(u) = P_\theta(U = u \mid y, s, \pi)$.  

The network architecture includes the following components. Input embedding layers process observations $y$ through embedding matrices that encode categorical variables (e.g., weather conditions, time periods) and normalize continuous ones. Signal embeddings similarly represent the signal space $\mathcal{S}$. The policy $\pi$ is encoded either through a learned embedding or direct parameterization of the policy function. Hidden layers with ReLU activations capture nonlinear interactions among observations, signals, and policies, while batch normalization and dropout provide regularization. The output layer applies a softmax activation to yield a valid probability distribution over the action space $\mathcal{U}$. For policy optimization, the predicted most likely action is given by  
\begin{equation}
u_\theta(y, s, \pi) = \arg\max_{u \in \mathcal{U}} f_\theta(y, s, \pi)(u).
\end{equation}

\subsection{Training Data and Objectives}

In our exposition, we consider that we have access to historical data
\begin{equation}
\mathcal{D} = \{(x_i, y_i, s_i, \pi_i, u_i)\}_{i=1}^N,~ N\in\mathbb{N},
\end{equation}
where $(x_i, y_i)$ are sampled from the environment, $s_i$ is generated by policy $\pi_i$, and $u_i$ is the observed receiver action.
The neural network is trained to minimize cross-entropy loss, i.e.,
\begin{equation}
L(\theta) = -\frac{1}{N} \sum_{i=1}^N \log f_\theta(y_i, s_i, \pi_i)(u_i) + \lambda \|\theta\|_2^2,
\label{eq:training_loss}
\end{equation}
where $\lambda \|\theta\|_2^2$ is $\ell_2$ regularization to prevent overfitting.

After training, the sender’s robust optimization problem can be written as  
\begin{equation}
\max_{\pi \in \Pi} \mathbb{E}_{X,Y,S}^\pi\left[\min_{u \in C_{1-\alpha}(Y,S,\pi)} r_s(X, u)\right],
\end{equation}
where $C_{1-\alpha}(y, s, \pi)$ denotes the conformal prediction set constructed in Section~\ref{sec:conformal_prediction} using the trained model $f_\theta$. This formulation enables policy optimization with finite-sample robustness guarantees derived from the conformal prediction framework.

\section{Conformal Prediction for Receiver Actions}
\label{sec:conformal_prediction}

Given the trained neural network $f_\theta$, we construct conformal prediction sets providing finite-sample guarantees for receiver actions. The key idea is to quantify uncertainty using nonconformity scores, which measure how unusual an action is given the model's predictions.


For discrete action spaces, we use the indicator-based nonconformity score
\begin{equation}
e_\theta(y, s, \pi, u) = \mathbb{I}\{u \neq u_\theta(y, s, \pi)\},
\label{eq:nonconformity_score}
\end{equation}
where $u_\theta(y, s, \pi) = \arg\max_{u} f_\theta(y, s, \pi)(u)$ is the predicted action.
This score equals zero when the action matches the prediction and one otherwise. This indicator-based score is particularly appropriate for classification tasks with discrete action spaces, as it directly measures prediction correctness. Alternative scores such as probability-based nonconformity $1 - f_\theta(y,s,\pi)(u)$ could also be used, with the choice depending on whether absolute prediction accuracy or confidence calibration is more important for the application.


Conformal prediction requires exchangeability of data points for valid coverage guarantees.

\begin{assumption}[Exchangeability]
\label{ass:exchangeability}
The data tuples $d_i := (X_i, Y_i, S_i, \Pi_i, U_i)$ for $i = 1, \dots, N + 1$ are exchangeable. That is, for any permutation $\sigma$ of $\{1, \ldots, N + 1\}$, the joint distribution satisfies
\begin{equation}
\mathbb{P}(d_1, \dots, d_{N+1}) = \mathbb{P}(d_{\sigma(1)}, \dots, d_{\sigma(N+1)}).
\end{equation}
\end{assumption}

This assumption holds when the data are i.i.d. from a fixed distribution or when they are collected from multiple policies in a randomized fashion. Exchangeability is required for the full tuples $(X_i, Y_i, S_i, \Pi_i, U_i)$, but not for the individual components. For example, the signals $S_i$ are not exchangeable on their own since they depend on the states $X_i$ through the policy, yet the joint tuples remain exchangeable when policies are randomly selected. This condition is weaker than independence and allows more flexible data collection procedures.

\begin{theorem}[Conformal Coverage Guarantee]
\label{thm:neural_exchangeability}
Let $\mathcal{D} = \{(x_i, y_i, s_i, \pi_i, u_i)\}_{i=1}^N$ denote the training data, and let $(x_{N+1}, y_{N+1}, s_{N+1}, \pi_{N+1}, u_{N+1})$ be a new test point. Suppose the neural network $f_{\hat{\theta}}$ is trained on $\mathcal{D}$, and define the conformal prediction set
\[
C_{1-\alpha}(y, s, \pi) = \{u \in \mathcal{U} : e_{\hat{\theta}}(y, s, \pi, u) \leq q_{1-\alpha}\},
\]
where $q_{1-\alpha} = \text{Quantile}_{1-\alpha}(\{e_{\hat{\theta}}(y_i, s_i, \pi_i, u_i)\}_{i=1}^N)$ is the $(1-\alpha)$-quantile of the nonconformity scores.  

Then, under Assumption~\ref{ass:exchangeability},
\begin{equation}
\mathbb{P}\{u_{N+1} \in C_{1-\alpha}(y_{N+1}, s_{N+1}, \pi_{N+1})\} \geq 1 - \alpha.
\end{equation}
\end{theorem}

\begin{proof}
The result follows from standard conformal prediction theory~\cite{vovk2005algorithmic}. 
By exchangeability of $(X_i, Y_i, S_i, \Pi_i, U_i)_{i=1}^{N+1}$, the nonconformity scores 
$\{e_{\hat{\theta}}(y_i, s_i, \pi_i, u_i)\}_{i=1}^{N+1}$ computed using the same function $e_{\hat{\theta}}$ 
are also exchangeable. Consequently, the rank of 
$e_{\hat{\theta}}(y_{N+1}, s_{N+1}, \pi_{N+1}, u_{N+1})$ among these $N+1$ scores is uniformly distributed 
on $\{1, \ldots, N+1\}$.  

Let $R$ denote this rank. The test point is included in the conformal set when its nonconformity score does not 
exceed the $(1-\alpha)$-quantile, i.e., when $R \leq \lceil (1-\alpha)(N+1) \rceil$. Since $R$ is uniformly distributed,  
\begin{equation}
\mathbb{P}\{R \leq \lceil (1-\alpha)(N+1) \rceil\} 
= \frac{\lceil (1-\alpha)(N+1) \rceil}{N+1} 
\geq 1 - \alpha.
\end{equation}
This probability is conditional on the training data $\mathcal{D}$ but holds for any realization of $\mathcal{D}$, 
yielding the unconditional guarantee. The event $\{R \leq \lceil (1-\alpha)(N+1) \rceil\}$ is equivalent to 
$\{u_{N+1} \in C_{1-\alpha}(y_{N+1}, s_{N+1}, \pi_{N+1})\}$, and the proof is complete.
\end{proof}

\subsection{Multi-Policy Learning}

When data are collected from multiple policies, we can construct a single conformal calibration valid across all policies in the training set.

We collect data from $K$ different policies as
\begin{equation}
\mathcal{D} = \bigcup_{k=1}^K \{(x_i^{(k)}, y_i^{(k)}, s_i^{(k)}, \pi_k, u_i^{(k)})\}_{i=1}^{N_k},
\end{equation}
where $N = \sum_{k=1}^K N_k$ is the total sample size. The neural network $f_\theta(y, s, \pi)$ is trained on pooled data, with the policy $\pi$ encoded as a network input through learned embeddings or direct parameterization. The data are split into training set $\mathcal{D}_{\text{train}}$ and calibration set $\mathcal{D}_{\text{cal}}$. Nonconformity scores are computed on the calibration set as $e_i = e_{\hat{\theta}}(y_i, s_i, \pi_i, u_i)$ for $i \in \mathcal{D}_{\text{cal}}$, and the threshold is set as $q_{1-\alpha} = \text{Quantile}_{1-\alpha}(\{e_i\})$.

For any policy $\pi$ whose behavior is represented in the training data (having sufficient data points), Theorem~\ref{thm:neural_exchangeability} ensures the coverage guarantee
\begin{equation}
\mathbb{P}\{U_{\text{new}} \in C_{1-\alpha}(Y_{\text{new}}, S_{\text{new}}, \pi)\} \geq 1 - \alpha.
\end{equation}
The coverage guarantee holds exactly for policies in this training set, while coverage for interpolated or extrapolated policies depends on the policy shift bounds established in Section~\ref{sec:theoretical_guarantees}.

\subsection{Single-Policy Learning with Policy Transfer}

In the more practical setting, we have data from only a single baseline policy $\hat{\pi}$. This scenario arises naturally when organizations have historical data from one operational policy but wish to optimize under alternative policies without incurring the cost of additional data collection.

The procedure consists of the following steps. First, train neural network $f_\theta(y, s, \pi)$ on data from policy $\hat{\pi}$, ensuring the network architecture can handle arbitrary policy inputs even though training uses only $\hat{\pi}$. Second, compute nonconformity scores: for each data point $i$, compute the predicted action as
\begin{equation}
\hat{u}_i = u_{\hat{\theta}}(y_i, s_i, \hat{\pi}) = \arg\max_{u \in \mathcal{U}} f_{\hat{\theta}}(y_i, s_i, \hat{\pi})(u),
\end{equation}
and define the nonconformity score (either using a simple indicator function $e_i = \mathbb{I}\{u_i \neq \hat{u}_i\}$ or a probability-based function $e_\theta(y, s, \pi, u) = 1 - f_\theta(y, s, \pi)(u)$). Third, compute the quantile threshold as $q_{1-\alpha} = \text{Quantile}_{1-\alpha}(\{e_i\}_{i=1}^N)$ using the empirical quantile from the training data. Fourth, construct policy-adaptive conformal sets: for any candidate policy $\pi$ and observation-signal pair $(y, s)$, define
\begin{equation}
C_{1-\alpha}(y, s, \pi) = \{u \in \mathcal{U} : e_{\hat{\theta}}(y, s, \pi, u) \leq q_{1-\alpha}\},
\end{equation}
where the predicted action $u_{\hat{\theta}}(y, s, \pi)$ is computed using the new policy $\pi$. Fifth, perform robust policy optimization.

This approach offers several benefits. It requires data from only one operationally feasible policy, reducing data collection costs. Conformal sets automatically adjust for different policies through the learned function $f_{\hat{\theta}}(y, s, \pi)$, which captures how policy changes affect receiver responses. The framework enables optimization over the entire policy space $\Pi$ while exploiting patterns learned from the baseline policy. However, coverage guarantees degrade when the new policy $\pi$ differs substantially from $\hat{\pi}$, as quantified by Theorem~\ref{thm:policy_shift_coverage} in Section~\ref{sec:theoretical_guarantees}.

\section{Theoretical Guarantees and Performance Bounds}
\label{sec:theoretical_guarantees}

This section establishes theoretical guarantees, including coverage under policy shift, neural network approximation, estimation error bounds, sample complexity, and robust optimization performance.

\subsection{Coverage under Policy Shift}

We provide coverage bounds when applying conformal sets constructed from data under policy $\hat{\pi}$ to predictions under a different policy $\pi$.
For two policies $\pi$ and $\hat{\pi}$, we define three shift measures that characterize the degradation in coverage guarantees.

The joint distribution shift measures how much the $(Y,S)$ distribution changes under different policies:
\begin{align}
\Delta_{\text{TV}}(\pi, \hat{\pi}) &= \text{TV}(P_{Y,S|\pi}, P_{Y,S|\hat{\pi}}) \\
&= \frac{1}{2}\sum_{y,s}|P_{Y,S|\pi}(y,s) - P_{Y,S|\hat{\pi}}(y,s)|. \nonumber
\end{align}
This measures whether the new policy induces a different distribution over observation-signal pairs, which affects which regions of the input space are evaluated.

The conditional mechanism shift measures whether the policy itself affects how receivers respond to signals:
\begin{equation}
\Delta_{\text{mech}}(\pi, \hat{\pi}) = \sup_{y,s} \text{TV}(P_{U|Y=y,S=s,\pi}, P_{U|Y=y,S=s,\hat{\pi}}).
\end{equation}
This captures whether receivers condition their actions on the policy itself, rather than only on the immediate observation-signal pair. For example, if receivers learn to anticipate policy patterns and adjust their responses accordingly, $\Delta_{\text{mech}}$ will be large.

The prediction calibration error measures how well nonconformity scores calibrate across policies:
\begin{align}
\Delta_{\text{cal}}(\pi, \hat{\pi}) = \big|\mathbb{E}_{Y,S,U}^\pi& [e_{\hat{\theta}}(Y,S,\pi,U)]\\
&- \mathbb{E}_{Y,S,U}^{\hat{\pi}}[e_{\hat{\theta}}(Y,S,\hat{\pi},U)] \big|. \nonumber
\end{align}
This measures whether the expected nonconformity score changes under the new policy, which affects the validity of using the quantile threshold $q_{1-\alpha}$ computed under $\hat{\pi}$.

\begin{theorem}[Coverage under Policy Shift]
\label{thm:policy_shift_coverage}
Consider data collected under a single fixed policy $\hat{\pi} \in \Pi$, 
$\mathcal{D} = \{(x_i, y_i, s_i, u_i)\}_{i=1}^N$, where $s_i \sim \hat{\pi}(\cdot|x_i)$ and 
$u_i \sim P_{U|Y,S,\Pi}(\cdot|y_i, s_i, \hat{\pi})$. 
Let $f_{\hat{\theta}}$ be trained on $\mathcal{D}$, and let $C_{1-\alpha}(y, s, \pi)$ denote the conformal prediction 
set constructed using nonconformity scores from $\mathcal{D}$.

\textbf{Part 1 (Exact coverage for the data-generating policy):}  
For the policy $\hat{\pi}$ generating the data,
\begin{equation}
\mathbb{P}\{U_{\text{new}} \in C_{1-\alpha}(Y_{\text{new}}, S_{\text{new}}, \hat{\pi}) \mid \hat{\pi}\} \geq 1 - \alpha.
\end{equation}

\textbf{Part 2 (Coverage bounds under policy shift):}  
For any alternative policy $\pi \in \Pi$, the coverage probability satisfies
\begin{align}
\mathbb{P}&\{U_{\text{new}} \in C_{1-\alpha}(Y_{\text{new}}, S_{\text{new}}, \pi) \mid \pi\} \nonumber\\
&\geq 1 - \alpha 
- 2\Delta_{\text{TV}}(\pi, \hat{\pi})
- \Delta_{\text{mech}}(\pi, \hat{\pi})
- \Delta_{\text{cal}}(\pi, \hat{\pi}),
\end{align}
where $\Delta_{\text{TV}}(\pi, \hat{\pi})$, $\Delta_{\text{mech}}(\pi, \hat{\pi})$, and $\Delta_{\text{cal}}(\pi, \hat{\pi})$ 
denote the total variation, mechanism shift, and calibration error terms, respectively.
\end{theorem}

\begin{proof}
\textbf{Part 1:} 
By Theorem~\ref{thm:neural_exchangeability}, under Assumption~\ref{ass:exchangeability}, the conformal set 
$C_{1-\alpha}(y, s, \hat{\pi})$ satisfies
\begin{equation}
\mathbb{P}\{U_{\text{new}} \in C_{1-\alpha}(Y_{\text{new}}, S_{\text{new}}, \hat{\pi}) \mid \hat{\pi}\} = 1 - \alpha.
\end{equation}

\textbf{Part 2:} 
Let $\mathcal{E}_\pi := \{U_{\text{new}} \in C_{1-\alpha}(Y_{\text{new}}, S_{\text{new}}, \pi)\}$ and 
$\mathcal{E}_{\hat{\pi}} := \{U_{\text{new}} \in C_{1-\alpha}(Y_{\text{new}}, S_{\text{new}}, \hat{\pi})\}$.  
Define the miscoverage probability $\varepsilon(\pi) := \mathbb{P}(\mathcal{E}_\pi^c \mid \pi)$.  
We decompose
\begin{align}
\varepsilon(\pi) 
&= \mathbb{P}\{e_{\hat{\theta}}(Y, S, \pi, U) > q_{1-\alpha}^{(\hat{\pi})} \mid \pi\} \nonumber\\
&= \mathbb{E}_{Y,S|\pi}\!\left[\int \mathbf{1}\{e_{\hat{\theta}}(y,s,\pi,u) > q_{1-\alpha}^{(\hat{\pi})}\}
\, P_{U|Y,S,\pi}(du)\right].
\label{eq:miscoverage_def}
\end{align}

We bound $\varepsilon(\pi) - \varepsilon(\hat{\pi})$ by three components.

\emph{(i) Joint distribution shift:}  
For any measurable $A \subseteq \mathcal{Y} \times \mathcal{S} \times \mathcal{U}$,  
\[
|\mathbb{P}(A \mid \pi) - \mathbb{P}(A \mid \hat{\pi})| 
\leq 2\Delta_{\mathrm{TV}}(\pi, \hat{\pi}),
\]
by the definition of the total variation distance
\(\Delta_{\mathrm{TV}}(\pi, \hat{\pi}) 
:= \tfrac{1}{2}\!\int |p_{Y,S}^\pi - p_{Y,S}^{\hat{\pi}}|\, d(y,s)\).
Hence
\begin{equation}
|\varepsilon(\pi) - \varepsilon(\hat{\pi})| 
\leq 2\Delta_{\mathrm{TV}}(\pi, \hat{\pi}) + R_1,
\label{eq:joint_bound}
\end{equation}
where $R_1$ captures residual conditional discrepancies.

\emph{(ii) Conditional mechanism shift:}  
For each $(y,s)$, define 
\(\delta_{\mathrm{mech}}(y,s)
:= \mathrm{TV}(P_{U|Y,S,\pi}(\cdot|y,s), P_{U|Y,S,\hat{\pi}}(\cdot|y,s))\).  
Then,
\begin{align}
R_1 
&\leq \mathbb{E}_{Y,S|\pi}[\delta_{\mathrm{mech}}(Y,S)] 
\leq \Delta_{\mathrm{mech}}(\pi, \hat{\pi}),
\label{eq:mech_bound}
\end{align}
by definition of $\Delta_{\mathrm{mech}}$.

\emph{(iii) Calibration error:}  
Let $q_{1-\alpha}^{(\pi)}$ and $q_{1-\alpha}^{(\hat{\pi})}$ be the $(1-\alpha)$-quantiles of 
$e_{\hat{\theta}}(Y,S,\pi,U)$ under $\pi$ and $\hat{\pi}$, respectively.  
Then, using the Lipschitz continuity of the cumulative distribution function,
\begin{align}
|\varepsilon(\pi) - \mathbb{P}\{e_{\hat{\theta}}(Y,S,\pi,U) &> q_{1-\alpha}^{(\pi)} \mid \pi\}| 
\leq |F_\pi(q_{1-\alpha}^{(\pi)}) \nonumber\\
&- F_\pi(q_{1-\alpha}^{(\hat{\pi})})| \leq \Delta_{\mathrm{cal}}(\pi, \hat{\pi}),
\label{eq:cal_bound}
\end{align}
where $\Delta_{\mathrm{cal}}$ bounds the deviation between calibrated and applied quantiles.

Combining \eqref{eq:joint_bound}–\eqref{eq:cal_bound} yields
\begin{equation}
\varepsilon(\pi) \leq \varepsilon(\hat{\pi}) 
+ 2\Delta_{\mathrm{TV}}(\pi, \hat{\pi})
+ \Delta_{\mathrm{mech}}(\pi, \hat{\pi})
+ \Delta_{\mathrm{cal}}(\pi, \hat{\pi}).
\end{equation}
Since $\varepsilon(\hat{\pi}) = \alpha$, it follows that
\begin{align}
&\mathbb{P}\{U_{\text{new}} \in C_{1-\alpha}(Y_{\text{new}}, S_{\text{new}}, \pi) \mid \pi\}
\geq 1 - \alpha 
- 2\Delta_{\mathrm{TV}}(\pi, \hat{\pi})\nonumber\\
&- \Delta_{\mathrm{mech}}(\pi, \hat{\pi})
- \Delta_{\mathrm{cal}}(\pi, \hat{\pi}).
\end{align}
\end{proof}

\subsection{Practical Guidance for Policy Transfer}

Theorem~\ref{thm:policy_shift_coverage} provides actionable guidance for policy transfer when receivers know and condition on the signaling policy. Since receivers update beliefs using $\pi(s|x)$ explicitly, the mechanism shift $\Delta_{\text{mech}}(\pi, \hat{\pi})$ is generally non-zero, which makes it difficult to measure exact coverage bounds in single-policy learning.

In practice, we recommend a two-stage approach. First, utilize the trained neural network to estimate shift measures for candidate policies without collecting new data. The joint distribution shift $\Delta_{\text{TV}}(\pi, \hat{\pi})$ can be computed exactly from policy definitions, while the mechanism shift $\Delta_{\text{mech}}$ can be approximated by evaluating $\sup_{y,s} \text{TV}(f_\theta(y,s,\pi), f_\theta(y,s,\hat{\pi}))$ using the trained model. Complement these estimates with uncertainty quantification to identify high-risk policies. Second, for the most promising candidates with significant uncertainty, collect data strategically from a small number of policies. Prioritize policies with high expected value of information: those combining strong predicted utility with high shift uncertainty. Retrain the neural network on pooled multi-policy data, which improves interpolation and provides reliable coverage guarantees. This approach substantially reduces data requirements compared to exhaustive policy evaluation while maintaining principled coverage bounds.

\subsection{Neural Network Performance Guarantees}

In this subsection, we establish approximation and estimation error bounds for neural network predictions, sample-complexity bounds for policy optimization, and robust utility guarantees.

\begin{assumption}[Loss Function and Network Properties]
\label{ass:nn_properties}
The cross-entropy loss function is Lipschitz continuous with constant $L$. The neural network class $\mathcal{F}_N$ has finite Rademacher complexity $\mathcal{R}_N(\mathcal{F}_N)$. The training data are i.i.d. from the true data-generating distribution.
\end{assumption}

\begin{theorem}[Neural Network Approximation and Estimation]
\label{thm:nn_approximation}
Let $f^*: \mathcal{Y} \times \mathcal{S} \times \Pi \to \Delta(\mathcal{U})$ denote the true conditional distribution of actions, and let $f_{\theta_{\text{best}}}$ denote the best approximation within the neural network class $\mathcal{F}_N$. Let $\hat{\theta}_N$ be the empirical risk minimizer over $N$ training samples.

Then under Assumption~\ref{ass:nn_properties}, the excess risk decomposes as
\begin{equation}
\mathbb{E}[\ell(U, f_{\hat{\theta}_N}(Y, S, \Pi))] - \mathbb{E}[\ell(U, f^*(Y, S, \Pi))] \leq \varepsilon_{\text{approx}} + \varepsilon_{\text{est}}
\end{equation}
where $\varepsilon_{\text{approx}} = \mathbb{E}[\ell(U, f_{\theta_{\text{best}}})] - \mathbb{E}[\ell(U, f^*)]$ and
\begin{equation}
\varepsilon_{\text{est}} \leq 2\mathcal{R}_N(\mathcal{F}_N) + 3L\sqrt{\frac{\log(2/\delta)}{2N}},
\end{equation}
with probability at least $1 - \delta$.
\end{theorem}

\begin{proof}
The proof follows from standard empirical risk minimization theory \cite{shalev2014understanding}. Decomposing the excess risk into approximation and estimation components, and applying generalization bounds based on Rademacher complexity \cite{bartlett2002rademacher}, yields the stated result via the union bound.
\end{proof}

The approximation error depends on network expressiveness and vanishes for sufficiently deep networks by universal approximation theorems \cite{hornik1989multilayer}. The estimation error decreases at a rate $O(1/\sqrt{N})$ but increases with model complexity through $\mathcal{R}_N(\mathcal{F}_N)$.

\begin{theorem}[Sample Complexity for Policy Optimization]
\label{thm:sample_complexity}
For $\varepsilon$-optimal policy selection with confidence $1 - \delta$, the required sample size satisfies
\begin{equation}
N = O\left(\frac{d \log(|\mathcal{U}||\mathcal{Y}||\mathcal{S}|/\varepsilon) + \log(1/\delta)}{\varepsilon^2}\right),
\end{equation}
where $d$ is the effective dimension of the neural network class.
\end{theorem}

\begin{proof}
From Theorem~\ref{thm:nn_approximation}, the estimation error scales as $O(\sqrt{d \log N / N})$ for networks with effective dimension $d$ using standard Rademacher complexity bounds \cite{bartlett2017spectrally}. For $\varepsilon$-optimal policy value, we require $\varepsilon_{\text{est}} \leq \varepsilon/C$ for some constant $C$ depending on problem parameters. Solving $\sqrt{d \log N / N} \leq \varepsilon/C$ and incorporating confidence parameter $\delta$ through union bounds over action and observation-signal spaces yields the stated complexity.
\end{proof}

\subsection{Robust Optimization Performance}

We now provide a finite-sample lower bound on sender utility under the robust policy.
Given conformal prediction sets for receiver actions, we formulate the sender's robust optimization problem as
\begin{equation}
\max_{\pi \in \Pi} \mathbb{E}_{X,Y,S}^\pi\left[\min_{u \in C_{1-\alpha}(Y,S,\pi)} r_s(X, u)\right].
\end{equation}

\begin{theorem}[Robust Utility Lower Bound]
\label{thm:robust_bound}
Let $\hat{\pi} \in \arg\max_{\pi \in \Pi} \mathbb{E}_{X,Y,S}^\pi[\min_{u \in C_{1-\alpha}(Y,S,\pi)} r_s(X, u)]$ be the solution to the robust optimization problem. Assume sender rewards satisfy $r_s(x, u) \in [m, M]$ for all $(x, u) \in \mathcal{X} \times \mathcal{U}$. Then,
\begin{align}
&\mathbb{E}_{X,Y,S, U_{\text{true}}}^{\hat{\pi}}[r_s(X, U_{\text{true}})] \nonumber\\ 
&\geq \mathbb{E}_{X,Y,S}^{\hat{\pi}}\left[\min_{u \in C_{1-\alpha}(Y,S,\hat{\pi})} r_s(X, u)\right] - \alpha(M - m),
\end{align}
with probability at least $1 - \alpha$ over the randomness in conformal set construction.
\end{theorem}

\begin{proof}
Define the coverage event $\mathcal{E} = \{U_{\text{true}} \in C_{1-\alpha}(Y, S, \hat{\pi})\}$. By the conformal prediction coverage guarantee (Theorem~\ref{thm:neural_exchangeability}), we have $\mathbb{P}(\mathcal{E}) \geq 1 - \alpha$.

We decompose the expectation by conditioning on the coverage event:
\begin{align} \label{eqn:expectation}
&\mathbb{E}[r_s(X, U_{\text{true}}) - \min_{u \in C_{1-\alpha}(Y,S,\hat{\pi})} r_s(X, u)] \nonumber\\
&= \mathbb{E}\left[\left(r_s(X, U_{\text{true}}) - \min_{u \in C_{1-\alpha}} r_s(X, u)\right) \mathbb{I}_{\mathcal{E}}\right] \nonumber\\
&\quad+ \mathbb{E}\left[\left(r_s(X, U_{\text{true}}) - \min_{u \in C_{1-\alpha}} r_s(X, u)\right) \mathbb{I}_{\mathcal{E}^c}\right],
\end{align}
where we suppress arguments for brevity.

On the coverage event $\mathcal{E}$, by definition of the conformal set, we have $U_{\text{true}} \in C_{1-\alpha}(Y, S, \hat{\pi})$, which implies
\begin{equation}
r_s(X, U_{\text{true}}) \geq \min_{u \in C_{1-\alpha}(Y,S,\hat{\pi})} r_s(X, u).
\end{equation}
Therefore, the first term in \eqref{eqn:expectation} is non-negative, i.e.,
\begin{equation}
\mathbb{E}[(r_s(X, U_{\text{true}}) - \min_{u \in C_{1-\alpha}} r_s(X, u)) \mathbb{I}_{\mathcal{E}}] \geq 0.
\end{equation}

On the complementary event $\mathcal{E}^c$, using the boundedness assumption $r_s(x, u) \in [m, M]$, we have
\begin{align}
&|r_s(X, U_{\text{true}}) - \min_{u \in C_{1-\alpha}} r_s(X, u)| \nonumber\\
&\leq \max_{u,u'} |r_s(X, u) - r_s(X, u')| \leq M - m.
\end{align}
Thus, the second term in \eqref{eqn:expectation}
is lower bounded by
\begin{align}
&\mathbb{E}\left[\left(r_s(X, U_{\text{true}}) - \min_{u \in C_{1-\alpha}} r_s(X, u)\right) \mathbb{I}_{\mathcal{E}^c}\right] \nonumber\\
&\geq -\mathbb{E}[(M - m)\mathbb{I}_{\mathcal{E}^c}] = -(M - m)\mathbb{P}(\mathcal{E}^c) \nonumber\\
&\geq -(M - m)\alpha.
\end{align}

Combining the lower bounds on both terms yields
\begin{equation}
\mathbb{E}[r_s(X, U_{\text{true}})] - \mathbb{E}[\min_{u \in C_{1-\alpha}(Y,S,\hat{\pi})} r_s(X, u)] \geq -\alpha(M - m),
\end{equation}
which rearranges to the stated bound.
\end{proof}

\begin{remark}[Tightness of the Bound]
Theorem~\ref{thm:robust_bound} shows that the robust approach provides a performance guarantee within $\alpha(M - m)$ of the true expected utility. This bound is tight in the worst case when miscoverage events consistently lead to worst-case reward differences, but can be substantially better in practice under the following conditions. When conformal sets are small due to high-confidence predictions, the minimum over the set is close to the true action's reward. When rewards are relatively uniform with a small range $ M-m$, the worst-case penalty is small even under miscoverage. When we choose a small $\alpha$ for tighter coverage, the bound improves linearly, though this increases conformal set sizes and may reduce robust utility.
\end{remark}

\section{Numerical Experiments}
\label{sec:experiments}
We validate our framework on a smart-grid safety problem where a central controller (sender) communicates with a local operator (receiver). The sender observes the true grid state $x \in \mathcal{X} = \{\text{stable (S), critical (C), unstable (U)}\}$ with prior $\mu_X = (0.50, 0.35, 0.15)$. The receiver observes local stress level $y \in \mathcal{Y} = \{\text{low }(\ell), \text{ nominal }(n), \text{ high }(h)\}$ through likelihood $\mu_{Y|X}$ and chooses action $u \in \mathcal{U} = \{\text{normal (N), curtail (C), shutdown (D)}\}$ after receiving signal $s \in \mathcal{S} = \{\text{low, med, high}\}$.

\begin{table*}[h]
\centering
\small
\caption{Likelihood and reward functions}
\label{tab:parameters}
\vspace{-2mm}
\begin{tabular}{c|ccc|c|ccc|c|ccc}
\hline
& \multicolumn{3}{c|}{$\mu_{Y|X}$} & & \multicolumn{3}{c|}{$r_r(x,u)$} & & \multicolumn{3}{c}{$r_s(x,u)$} \\
& $\ell$ & $n$ & $h$ & & N & C & D & & N & C & D \\
\hline
S & .70 & .25 & .05 & S & 20 & 6 & -20 & S & 8 & 4 & -50 \\
C & .15 & .60 & .25 & C & 10 & 5 & -5 & C & -100 & 1 & -20 \\
U & .05 & .25 & .70 & U & -100 & -10 & 30 & U & -800 & -50 & 10 \\
\hline
\end{tabular}
\vspace{-2mm}
\end{table*}

The observation likelihood and the utility functions are given in Table \ref{tab:parameters}. The observation likelihood $\mu_{Y|X}$ captures realistic correlations where stable states predominantly yield low observations, critical states yield nominal, and unstable states yield high. The receiver utility $r_r(x,u)$ balances operational cost against blackout avoidance, while sender utility $r_s(x,u)$ emphasizes system stability with catastrophic penalties ($r_s(U,N) = -800$) for failing to curtail unstable conditions. We model receiver behavior through approximate Bayesian updating and prior misspecification (25\% mean deviation from $\mu_X$).

The neural network has input dimension 15 (one-hot encoding for $y \in \mathcal{Y}$, $s \in \mathcal{S}$, and policy $\pi$), two hidden layers (128, 64 neurons with ReLU, batch normalization, dropout 0.3), and softmax output. Training uses cross-entropy loss with $\ell_2$ regularization ($\lambda = 0.001$), AdamW optimizer (learning rate $5 \times 10^{-3}$ with ReduceLROnPlateau), and early stopping (patience 30). For conformal prediction, we employ negative log-likelihood nonconformity score $e_\theta(y,s,\pi,u) = -\log(f_\theta(y,s,\pi)(u) + \epsilon)$ with Adaptive Prediction Sets construction.

\begin{figure*}
    \centering
    \includegraphics[width=0.45\linewidth]{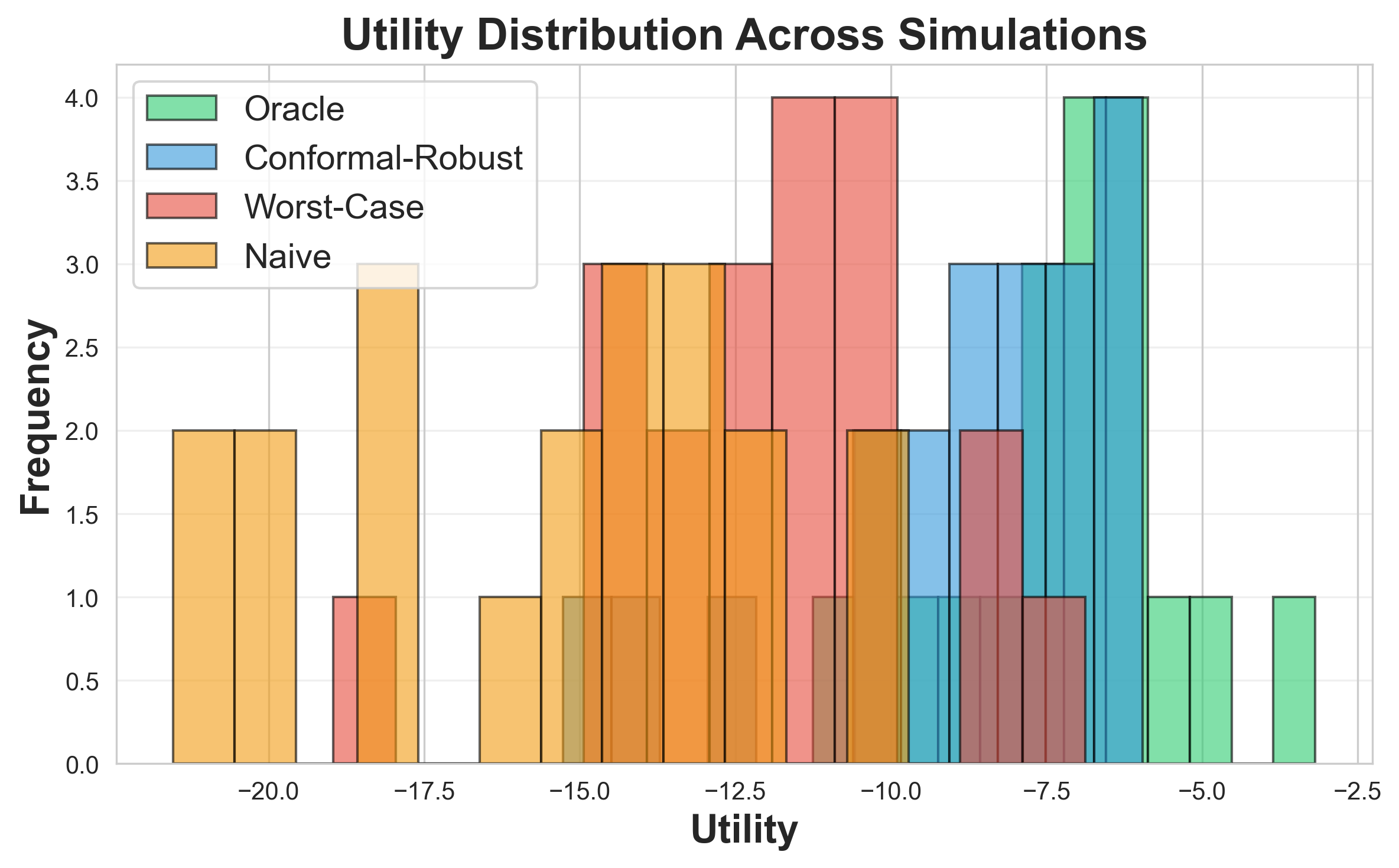}
    \includegraphics[width=0.45\linewidth]{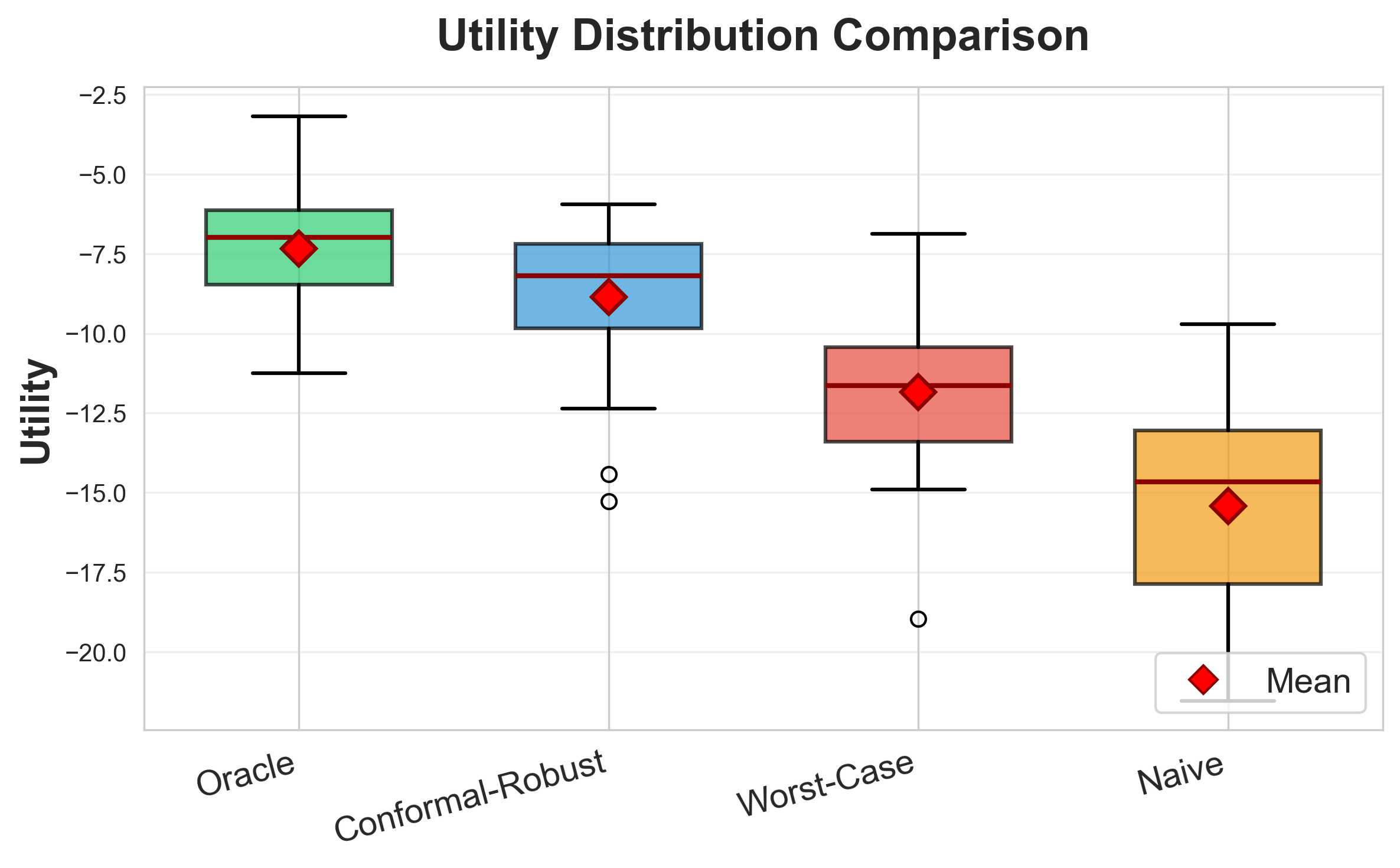}
    \caption{Sender utility across 20 simulations for four methods: (left) frequency distribution showing the spread of utilities;  (right) box plot of utility distributions with median, quartiles, and outliers.}
    \label{fig:utilities}
\end{figure*}


We compare four approaches in the experiments. \textbf{Oracle} optimizes assuming perfect knowledge of the receiver's belief-formation and decision process and thus provides an upper bound on performance. \textbf{Conformal-Robust} is our proposed single-policy procedure (target coverage $1-\alpha=0.90$). \textbf{Worst-Case} corresponds to classical robust optimization, and \textbf{Naive} denotes a baseline that optimizes standard Bayesian persuasion under the assumption of perfectly Bayesian receivers with the true prior (ignoring private observations and bounded rationality).

Figure \ref{fig:utilities} presents the utility distribution across 20 simulations. We evaluated each method with $500$ test samples, and performance were in the order of Oracle ($-7.34 \pm 1.88$) $>$ Conformal-Robust ($\mathbf{-8.85 \pm 2.54}$) $>$ Worst-Case ($-11.84 \pm 2.61$) $>$ Naive ($-15.42 \pm 3.43$), where our method achieves 80\% of oracle utility while naive achieves only 48\%. For the selected Conformal-Robust policy, conformal calibration computed on held-out calibration data (split from the single-policy dataset) produced a baseline empirical coverage of \(\mathbf{88.2}\%\). Performing the policy-specific re-calibration on samples generated under the selected policy raised empirical coverage to \(\mathbf{94.8}\%\).

\begin{figure}
    \centering
    \includegraphics[width=\linewidth]{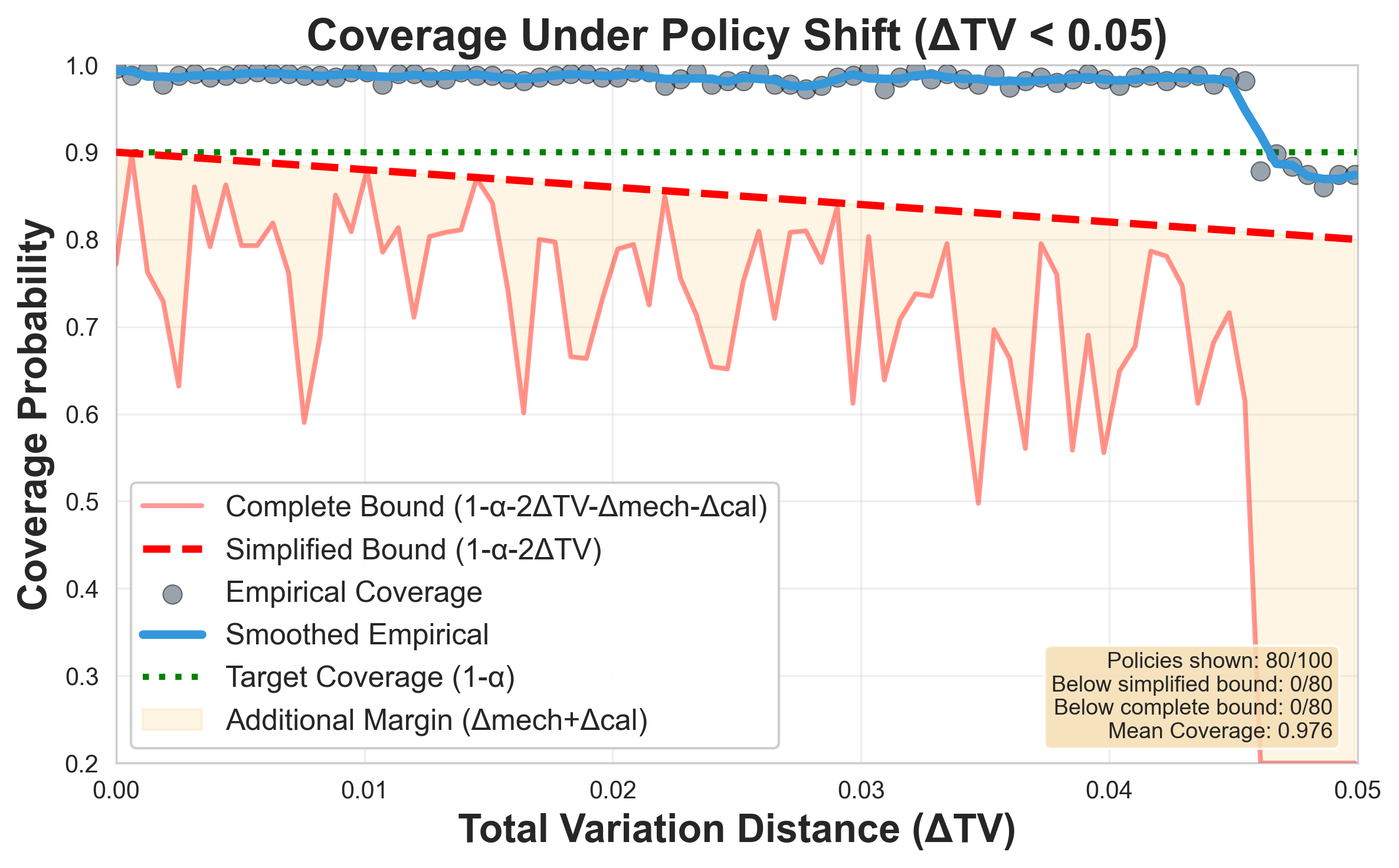}
    \caption{Empirical coverage under policy shift.}
    \label{fig:coverage}
\end{figure}

To validate Theorem~\ref{thm:policy_shift_coverage}, we examined coverage under controlled policy perturbations. Using a sequence of candidate policies with varying total-variation distances \(\Delta_{\text{TV}}(\pi,\hat\pi)\) (up to 0.05), we measured (i) empirical coverage, (ii) conditional mechanism shift, and (iii) calibration error.
As illustrated in Fig. \ref{fig:coverage}, empirical coverage generally met or exceeded the nominal \(90\%\) level for small $\Delta_{\text{TV}}$.
Meanwhile, we observed that the theoretical bound can be quite conservative and exhibit variability driven primarily by the mechanism-shift and calibration-error terms. Thus, it is recommended to recalibrate the conformal set whenever the chosen policy departs substantially from the baseline policy.

\section{Conclusion}
\label{sec:conclusion}

In this paper, we introduced a distributionally robust framework for Bayesian persuasion that remains effective when the sender faces uncertainty about the receiver’s belief formation. By combining neural network–based behavioral modeling with conformal prediction, the proposed approach constructs finite-sample valid uncertainty sets for receiver actions and enables robust policy optimization without explicit distributional assumptions. 

We established theoretical guarantees, including exact coverage under the data-generating policy, coverage degradation bounds under policy shift, neural network approximation and estimation error bounds, and a finite-sample lower bound on sender utility. Numerical experiments in smart-grid demand-response scenarios demonstrated the practicality and robustness of the framework in settings with private receiver information and behavioral heterogeneity. 

A potential direction for future research includes extending this approach to dynamic multi-stage persuasion problems, multi-agent interactions, and online adaptive schemes that refine uncertainty sets as new data become available.

\bibliographystyle{IEEEtran}
\bibliography{IDS,BP}

\end{document}